\documentclass[
  twoside,
  twocolumn,
  prb,
  preprintnumbers,
  floatfix,
  showpacs,
  superscriptaddress,
  citeautoscript,
]{revtex4-1}

\usepackage{amsmath}
\usepackage{amssymb}
\usepackage{color}
\usepackage{graphicx}
\usepackage{tabularx}
\usepackage{subfigure}
\usepackage{dcolumn}
\usepackage{times}
\newcolumntype{Y}{D..{4.2}}

\newcommand{\sect}[1]{Sect.~\ref{#1}}
\newcommand{\fig}[1]{Fig.~\ref{#1}}
\newcommand{\Fig}[1]{Figure~\ref{#1}}
\newcommand{\eq}[1]{Eq.~(\ref{#1})}
\newcommand{\Eq}[1]{Equation~(\ref{#1})}
\newcommand{\tab}[1]{Table~\ref{#1}}

\renewcommand{\vec}[1]{\ensuremath\boldsymbol{#1}}
\renewcommand{\epsilon}[0]{\varepsilon}

\newcommand{\myscale}{0.65}

\begin{document}

\title{
  A variational polaron self-interaction corrected \\
  total-energy functional for charge excitations in insulators 
}

\author{Babak Sadigh}
\email{sadigh1@llnl.gov}
\affiliation{
  Lawrence Livermore National Laboratory,
  Physical and Life Sciences Directorate,
  Livermore, California, 94550, USA
}
\author{Paul Erhart}
\affiliation{
  Chalmers University of Technology,
  Department of Applied Physics,
  S-41296 Gothenburg, Sweden
}
\author{Daniel {\AA}berg}
\affiliation{
  Lawrence Livermore National Laboratory,
  Physical and Life Sciences Directorate,
  Livermore, California, 94550, USA
}

\date{\today}

\begin{abstract}
We conduct a detailed investigation of the polaron self-interaction (pSI) error in standard approximations to the exchange-correlation (XC) functional within density-functional theory (DFT). The pSI leads to delocalization error in the polaron wave function and energy, as calculated from the Kohn-Sham (KS) potential in the native charge state of the polaron. This constitutes the origin of the systematic failure of DFT to describe polaron formation in band insulators. It is shown that the delocalization error in these systems is, however, largely absent in the KS potential of the closed-shell neutral charge state. This leads to a modification of the DFT total-energy functional that corrects the pSI in the XC functional. The resulting pSIC-DFT method constitutes an accurate parameter-free {\it ab initio} methodology for calculating polaron properties in insulators at a computational cost that is orders of magnitude smaller than hybrid XC functionals. Unlike approaches that rely on parametrized localized potentials such as DFT+$U$, the pSIC-DFT method properly captures both site and bond-centered polaron configurations. This is demonstrated by studying formation and migration of self-trapped holes in alkali halides (bond-centered) as well as self-trapped electrons in an elpasolite compound (site-centered). The pSIC-DFT approach consistently reproduces the results obtained by hybrid XC functionals parametrized by DFT+$G_0W_0$ calculations. Finally, we generalize the pSIC approach to hybrid functionals, and show that in stark contrast to conventional hybrid calculations of polaron energies, the pSIC-hybrid method is insensitive to the parametrization of the hybrid XC functional. On this basis, we further rationalize the success of the pSIC-DFT approach. 
\end{abstract}

\maketitle

\section{Introduction}
\label{sec:intro}

Polarons form in many wide-band gap materials as a result of charge carrier localization due to strong electron-phonon coupling. Their formation and transport are crucial to understanding important quantum mechanical processes occurring in electrochemical devices such as batteries \cite{MaxZhoCed06} and scintillation detectors \cite{WilSon90}.  Conventional approximations to the exchange and correlation (XC) potential within density-functional theory (DFT), such as the local density approximation (LDA) and generalized gradient approximations (GGA) \cite{GavSusShl03}, fail qualitatively to describe polaron formation. Hybrid functionals, which combine exact exchange (EX) \cite{PerErnBur96,HeyScuErn03} with LDA/GGA to correct for the band gaps of insulators, are capable of predicting polaron formation but are computationally orders of magnitude more expensive than DFT.

For small polarons localized at specific atomic sites, which is the case in in many oxides, \cite{LanZun09, MorWat09, KeaScaMor12, ErhKleAbe14} an orbital-dependent local potential in the spirit of the DFT+$U$ approach \cite{AniZaaAnd91} can be added to localize the excess charge and stabilize the polaron. While this enables one to significantly reduce the computational cost for calculating polaron energies, the approach is not viable if the polaronic state involves multiple atomic sites and/or large atomic displacements. This is unfortunately a very common occurrence, {\it e.g.} in halides, where polaron formation proceeds via dimerization (bond-centered polarons, so-called $V_K$-centers\cite{Kan55}) or in the case of polaron migration. Furthermore, both hybrid-DFT and local potential methods depend on adjustable parameters. The fraction of EX or the strength of the local potential are usually determined {\it a priori} for a reference configuration and then kept fixed throughout configurational changes. In summary, numerous complications have limited the systematic investigation of polarons and their transport from first principles. 

The present work builds on two ideas: (i) the LDA/GGA failure to predict polaron formation is mainly due to the self-interaction (SI) of the localizing carrier \cite{LanZun09,ZawJacRos11,ZawRosWed11} and (ii) while LDA and GGA severely underestimate the band gaps of insulators, they accurately predict band edge variations, induced by configurational changes \cite{GygBal89,SadErhAbe11}. Based on these considerations, we propose a simple modification to the DFT total-energy functional, the {\em polaron self-interaction correction} (pSIC), that accounts for the spurious polaron SI without explicit addition of EX, and without an implicit limitation to site-centered local potentials. The resulting pSIC-DFT functional is parameter-free, can be used for {\it {ab initio}} molecular-dynamics simulations of hole as well as electron polarons in arbitrary band insulators with an accuracy that is on par with the best hybrid-DFT parametrizations, at a computational cost comparable with standard DFT. Its strength and versatility is demonstrated by studying formation and migration of hole-polarons in the alkali halide family of compounds, and electron polarons in Cs$_2$LiYCl$_6$ (CLYC), which is an elpasolite compound that has received much attention in recent years due to its potential as a scintillator material.

It is important to point out that the polaron self-interaction correction can be applied to any XC functional. Naturally, when applied to hybrid functionals, the pSIC method does not decrease the computational cost. However, also in these cases, pSIC leads to a systematic improvement and significantly reduces the sensitivity of the results to hybrid parametrization.
This lends further credibility to the pSIC method as a general formalism applicable to polarons in crystalline and disordered insulators as well as molecular systems, whenever the removal or addition of an electron from/to a frontier orbital (highest occupied or lowest unoccupied quasi-particle state) drives a structural change. 

This paper is organized as follows. The pSIC method is derived in \sect{sec:derivation}. Section~\ref{sec:results} presents a comprehensive investigation of the accuracy of the pSIC method in comparison with the hybrid technique. In particular, formation and migration of hole polarons in alkali halides, and formation of electron polarons in CLYC are studied. Section~\ref{sec:discussion} summarizes and discusses the findings of the present work. Finally, computational details are given in the Appendix.

\section{The pSIC method}
\label{sec:derivation}

In this section, we motivate and develop the pSIC method by revisiting the failure of standard DFT functionals for hole self-trapping in the alkali halide compound NaI \cite{GavSusShl03}. It is an ionic compound in the rocksalt crystal structure with an equilibrium lattice constant of 6.48\,\AA \cite{Cor92}. The crystal consists of two interpenetrating fcc sublattices, each consisting of either Na or I ions only.  The ionic bonding in NaI leads to an insulating ground-state electronic structure with a band gap of 5.9\,eV \cite{TeeBal67, BroGahFuj70, ErhSchSad14}. The valence band is dominated by hybridized iodine $p$-orbitals, while the conduction band consists mainly of sodium $s$-orbitals. When a hole is introduced in the valence band, it mediates covalent bonding between nearest neighbor iodine ions leading to dimerization; this implies hole localization and thus the formation of a polaron\cite{Kan55}. In the following section, the pathway to polaron formation in NaI is studied from various angles using DFT as well as hybrid XC functionals.

\subsection{Basic considerations}
\label{subsec:basic}

\begin{figure}
  \centering
  \includegraphics[scale=\myscale]{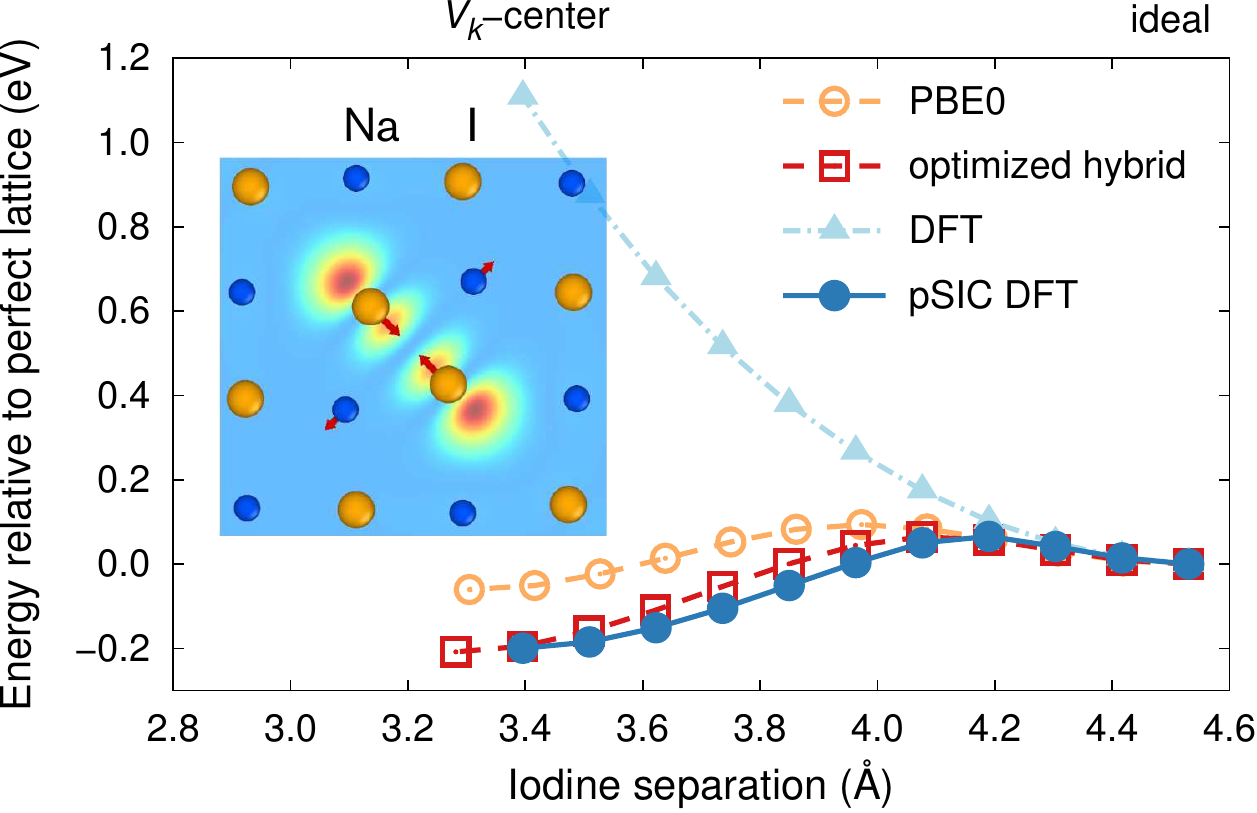}
  \caption{
    Energy as a function of the two iodine ions that form the core of the $V_K$-center in NaI. The ``optimized hybrid'' has an EX fraction of $0.31$. In agreement with the hybrid functionals pSIC-DFT yields a stable polaron configuration. Uncorrected DFT qualitatively fails to reproduce self-trapping.
  }
  \label{fig:self_trapping_path}
\end{figure}

When a hole is introduced in the valence band of NaI, it induces a substantial lattice distortion, causing a pair of nearest-neighbor I$^-$ ions to move along $\left<110\right>$ toward each other. Their separation is reduced from 4.5\,\AA\ in the perfect crystal to about 3.3\,\AA, effectively leading to the formation of an anion dimer I$_2^-$ (a $V_K$-center). The dimerization is accompanied by displacements of the surrounding atoms that help to accommodate the lattice strain. \Fig{fig:self_trapping_path} shows the energy along a particular pathway to self-trapping calculated from (i) DFT based on the PBE XC functional \cite{PerBurErn96}, (ii) the PBE0 hybrid-EX functional with 0.25 fraction of EX \cite{PerErnBur96}, and (iii) an optimized hybrid-EX functional in which the fraction of EX ($0.31$) has been chosen to reproduce the band gap of NaI calculated with PBE+$G_0W_0$. Further computational details are listed in the Appendix. \Fig{fig:self_trapping_path} clearly shows that when a hole charge is present in the system, the PBE XC functional predicts the undistorted lattice to be the lowest energy configuration. According to PBE, the hole in the valence shell is thus a homogeneously distributed delocalized carrier. On the other hand, \fig{fig:self_trapping_path} also demonstrates that hybrid functionals, which include some fraction of EX, allow the system to lower its energy via a lattice distortion that couples to the hole charge and triggers its localization. Hybrid functionals are thus capable of stabilizing hole polarons in NaI. From \fig{fig:self_trapping_path}, it is also evident that the polaron binding energy sensitively depends on the choice of the EX fraction in the hybrid functional.

\begin{figure*}
  \centering
  \includegraphics[width=0.95\linewidth]{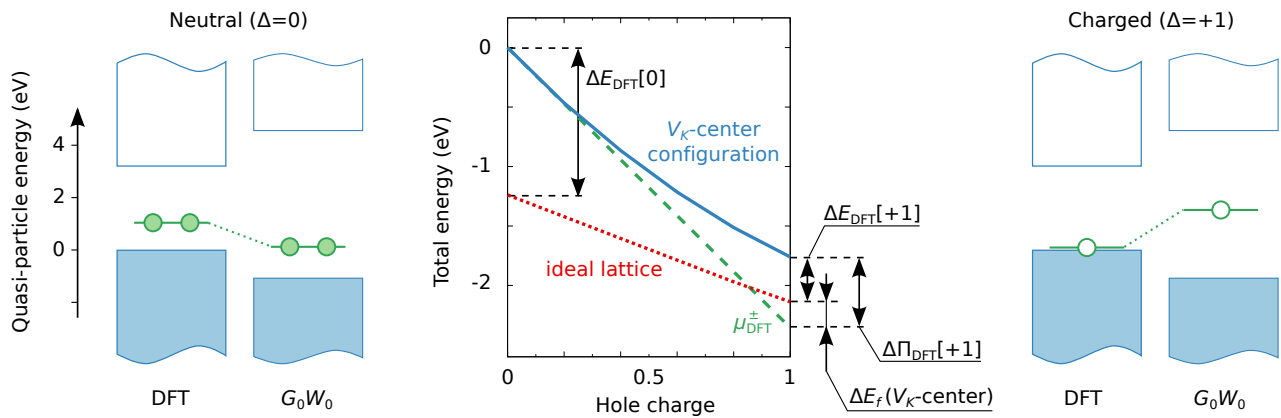}
  \caption{
     The center panel shows the variation of the total energy of the perfect lattice (dotted red line) and the $V_K$-center (solid blue line) configurations with fractional hole charge.  The SI can be corrected by removing the curvature of the solid blue line as illustrated by the dashed green line. The offset between the uncorrected (solid blue) and SI corrected in the limit $\Delta=+1$ then corresponds to the term $\Pi_{\text{DFT}}[\Delta=+1]$ in \eq{eq:sic_pi}. The left and right panels show side-by-side views of quasi-particle energies calculated within DFT and $G_0W_0$ for neutral and charged, respectively, 64-atom supercells of NaI containing a $V_K$-center.
  }
  \label{fig:pSIC}
\end{figure*}

\begin{figure}
  \centering
  \includegraphics[scale=\myscale]{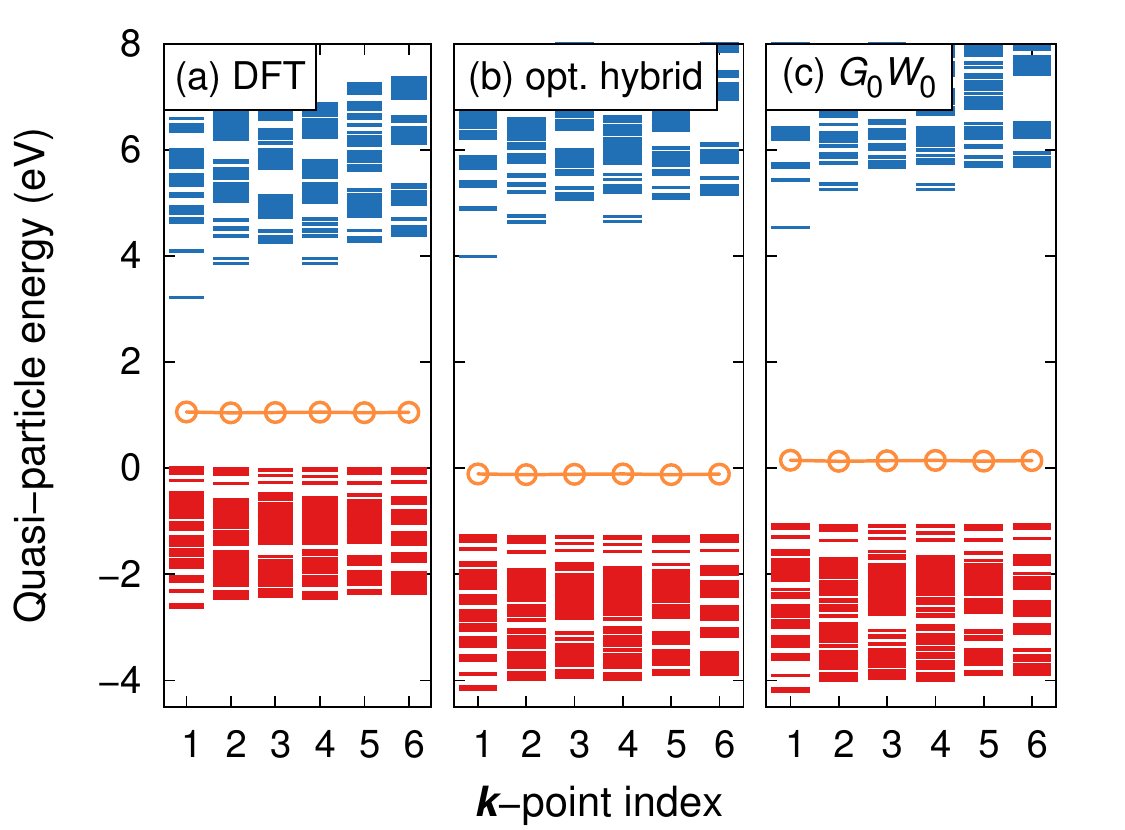}
  \caption{
   Energy levels for the $V_K$-center configuration calculated using DFT, optimized hybrid, and $G_0W_0$ (based on PBE0 calculations) for NaI. Red and blue lines represent occupied and unoccupied levels, respectively. The polaronic level is shown by orange lines and open circles.
  }
  \label{fig:NaI}
\end{figure}

It is important to mention here that in the absence of charge excitations, DFT predicts accurate ground state (GS) energies as well as phonon spectra for NaI. Furthermore, \fig{fig:pSIC} (left panel) shows that in the neutral charge state, PBE and $G_0W_0$ quasiparticle (QP) energies for a supercell containing atomic displacements of a $V_K$-center, are in good agreement with each other apart from a rigid relative shift of the valence and the conduction bands, the so-called scissor shift that compensates for the absence of derivative discontinuity in the PBE functional. A more detailed comparison between the QP spectra from PBE, optimized hybrid, and $G_0W_0$ is given in \fig{fig:NaI}. The close agreement of the QP energies implies that for the neutral charge state, the PBE-KS potential is a good approximation to the optimized effective potential (OEP) obtained from the exact XC functional.\footnote{
  Note that the eigenvalues of the exact OEP are not equivalent to the derivative of the energy functional with respect to the electron occupations. The difference makes up the so-called derivative discontinuity, see Refs.~\onlinecite{ShaSch83, *PerLev83}.
}
As a result, for {\em this} charge state PBE accurately predicts both the GS energy of the $V_K$-center, as well as the QP energy of the hole state relative to the valence band maximum (VBM). In contrast, \fig{fig:pSIC} (right panel) shows that in the charged state, the unoccupied hole level virtually overlaps with the VBM, which implies a delocalized hole character. This explains why PBE predicts no binding for the $V_K$-center in \fig{fig:self_trapping_path}. 

Before embarking on a formal discussion, it is beneficial to inspect the failure of DFT for the charged $V_K$-center in NaI from the perspective of the SI error of the localized hole. This is illustrated in the center panel of \Fig{fig:pSIC}, where the deviation of the DFT energy from linearity as a function of the fractional hole charge in the system is shown for (i) the perfect crystal and (ii) the dimerized $V_K$-center configuration. It is apparent that the delocalized hole carrier in the perfect crystal (with negligible SI) exhibits very little deviation from linearity while significant non-linearity is observed in the energetics of the localized polaron in the $V_K$-center configuration. In accord with recent literature \cite{MorCohYan08, LanZun09, MorWat09, DabFerPoi10, ZawRosWed11, ZawJacRos11, KeaScaMor12, DabFerMar14, ErhKleAbe14}, we ascribe this nonlinearity to the DFT SI error for the localized polaron. Note that the solid blue line in \fig{fig:pSIC} (center panel) curves in such a way as to raise the energy of the charged polaronic configuration relative to that of the ideal lattice, and thus prevents binding. The effect of the SI error on the stability of polarons within DFT has also been discussed in Ref.~\onlinecite{ZawJacRos11}. 

\subsection{The pSIC-DFT total energy functional}

We now formalize the above discussion with the goal to derive a robust method for structural relaxations as well as {\it ab initio} molecular dynamics simulations of polarons in general insulating systems with a computational efficiency comparable to LDA/GGA, which does not rely on {\it a priori} knowledge or guesswork. Consider an extended many-electron system in a periodic supercell containing $N_p$ ions, at a reference electronic configuration, say, the neutral ground state. In the following, we denote the number of electrons in this reference charge state by $N_e$. Let $E_{\text {DFT}}(N_e\pm \Delta;\vec{R})$ denote the DFT energy of this system as a function of the fraction $\Delta$ of excess electrons/holes. The ion positions are specified by the $3N_p$-dimensional vector $\vec{R}$, where $N_p$ is the number of ions in the system. We break down the contributions to the energy at arbitrary $\Delta$ as follows
\begin{align}
  \label{eq:sic_pi}
  E_{\text{DFT}}[N_e\pm \Delta;\vec{R}]
  =&
  E_{\text{DFT}}[N_e;\vec{R}]
  \pm \Delta\cdot\mu_{\text{DFT}}^{\pm}[\vec{R}]
   \nonumber\\
  &\quad + \Pi_{\text{DFT}}[\pm \Delta;\vec{R}],
\end{align}
where $\mu_{\text{DFT}}^{\pm}$ is the electron chemical potential, defined as 
\begin{align}
  \label{eq:mu}
\mu_{\text{DFT}}^{\pm}[\vec{R}]= \lim_{\eta\rightarrow 0}\frac{\partial E_{\text{DFT}}}{\partial N}[N_e\pm\eta;\vec{R}],
\end{align}
and $\Pi_{\text{DFT}}$ quantifies the deviation of $E_{\text{DFT}}$ from linearity and thus can be used as a measure of the polaron SI. \Fig{fig:pSIC} (center panel) shows that the $V_K$-center configuration becomes stabilized for $\Delta = 1$ if the term $\Pi_{\text{DFT}}$ (the polaron SI) is eliminated from the above equation.

At first sight, this seems to counter the well-known fact that the electron chemical potentials $\mu_{\text{DFT}}^{\pm}$, calculated within DFT, are inaccurate predictors of the highest occupied (HOMO) and lowest unoccupied molecular (LUMO) levels in molecules and of band gaps in solids. The curvatures of LDA/GGA functionals resulting from their SI are, however, such that they correct the values of the differential quantities $\mu_{\text{DFT}}^{\pm}$ for the HOMO/LUMO levels in small molecules. This constitutes the basis for the so-called delta-SCF method, which has been quite successful in molecular applications. It is however, a precarious practice to correct for the lack of derivative discontinuity in LDA/GGA XC functionals \cite{HybLou85, GodSchSha86} by their SI error. In particular, $\Pi_{\text{DFT}}[\pm \Delta;\vec{R}]$ can have a spurious dependence on the ionic configuration $\vec{R}$ and thus lead to wrong structural predictions. This can most dramatically be demonstrated in insulating solids such as NaI, where the perfect crystal structure has negligible SI while $\Pi_{\text{DFT}}$ is quite substantial for the self-trapped polaron. The strong configuration dependence of $\Pi_{\text{DFT}}$ leads to the failure to predict polaron formation in these systems. A systematic correction should rather correct $\mu_{\text{DFT}}^{\pm}$ by adding the contribution from the derivative discontinuity of the exact exchange-correlation functional, which is missing in LDA/GGA. Hence we propose the following polaron SI corrected (pSIC) total energy functional for a system with an added electron/hole:
\begin{align}
  E_{\text{pSIC}}^{\pm}[\vec{R}] = E_{\text{DFT}}[N_e;\vec{R}] \pm \mu_{\text{DFT}}^{\pm}[\vec{R}] \pm \Delta_\text{XC}^{\pm}[\vec{R}],
  \label{eq:final0}
\end{align}
where $\Delta_\text{XC}^{\pm}$ is the missing contribution of the derivative discontinuity of the exchange-correlation functional to the chemical potentials \cite{ShaSch83, *PerLev83}. It can be formally expressed as
\begin{align}
\Delta_\text{XC}^{\pm}[\vec{R}] = \lim_{\eta\rightarrow 0}\frac{\partial E^{\text{exact}}}{\partial N}[N_e\pm\eta;\vec{R}] - \epsilon^\text{exact-OEP}_{\pm}[N_e;\vec{R}],
\end{align}
where $\epsilon^\text{exact-OEP}_+$ and $\epsilon^\text{exact-OEP}_-$ denote the lowest unoccupied and the highest occupied eigenvalues of the exact OEP potential in the reference charge state, respectively. Note that for the $E_{\text{pSIC}}^{\pm}$ functional in \eq{eq:final0} to be accurate, it is required that 
\begin{align}
E_{\text{DFT}}[N_e;\vec{R}] &\approx E_{\text{exact}}[N_e;\vec{R}]+\text{constant}\\
\mu_\text{DFT}^{\pm}[\vec{R}]&\approx\epsilon^\text{exact-OEP}_{\pm}[N_e;\vec{R}] + \text{constant}.
\end{align}
The above is guaranteed, if for the reference charge state, the KS potential obtained from approximate XC functionals used in practice closely resembles the exact OEP. We have shown in \sect{subsec:basic} that this is indeed the case for the closed-shell neutral charge state of NaI. In fact, it is reasonable to argue that for defect-free band insulators in the closed-shell neutral charge state, common approximate XC functionals are likely to produce reasonably accurate KS potentials. 

The derivative discontinuity $\Delta^{\pm}_\text{XC}$, can be calculated accurately via many-body perturbation theory, {\it e.g.}, in the $GW$ approximation\cite{Hed65, *HedLun70, *AulJonWil00}. In this formalism,  $\Delta_\text{XC}^{\pm}$ correspond to the so-called scissor shifts, which constitute the difference between the electron addition/removal energies calculated using DFT+$G_0W_0$ and their respective DFT values. Furthermore, it has been shown by Baldereschi, Gygi and Fiorentini \cite{GygBal89, FioBal95} that $\Delta_\text{XC}^{\pm}$ is dominated by the long-range components of the static Coulomb-hole plus screened exchange (COHSEX) contributions. Therefore $\Delta_\text{XC}^{\pm}$ varies weakly with localized lattice distortions such as those typically associated with polarons, so long as the DFT produces accurate KS potentials for the ionic configurations of interest. Hence, we can simplify \eq{eq:final0} significantly by neglecting the dependence of $\Delta_\text{XC}^{\pm}$ on $\vec{R}$. Next, we observe that with regard to defects in general and polarons in particular, one is interested in energy differences. For example, for polarons in insulators, all energies are calculated relative to the undistorted crystal, in which the excess charge is delocalized. The detailed expressions for calculating polaron binding energies can be found in \sect{sec:binding} of the Appendix. Since we are only interested in energy differences, it is possible to move the zero of energy in such a way as to remove the contribution from the derivative discontinuity in the total-energy functional \eq{eq:final0}. We thus arrive at the expression
\begin{align}
\Delta E_{\text{pSIC-DFT}}^{\pm}[\vec{R}] = \Delta E_{\text{DFT}}[N_e;\vec{R}] \pm \Delta \mu_{\text{DFT}}^{\pm}[\vec{R}].
  \label{eq:final}
\end{align}
The notation $\Delta E$ and $\Delta\mu$ is used above to highlight the fact that we are only interested in relative energies.

\subsection{Generalization to pSIC-hybrid functionals}
\label{eq:pSIC-hyb}

\begin{figure}
  \centering
  \includegraphics[scale=\myscale]{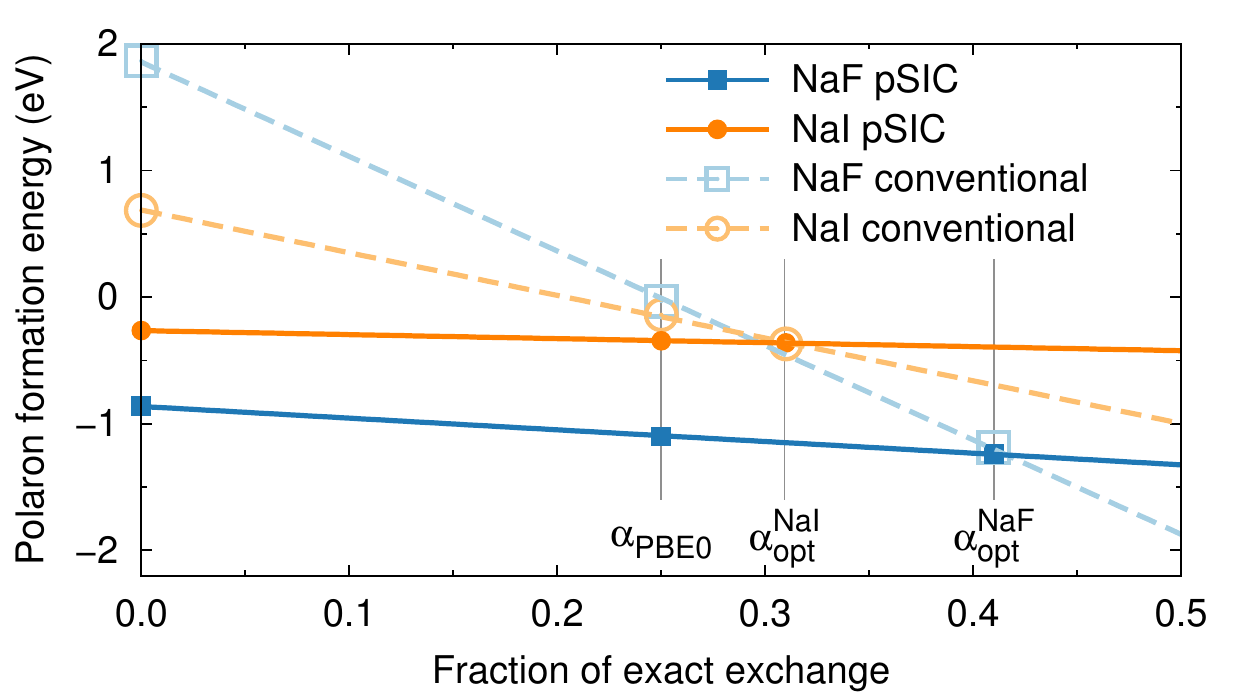}
  \caption{
    Variation of the calculated polaron formation energies of NaI and NaF with the mixing parameter of the hybrid XC functional. Note that for the conventional hybrid-XC functionals the polaron formation energy is strongly dependent on the mixing parameter. This situation is alleviated almost entirely by using the pSIC-DFT/pSIC-hybrid approach. The deviation between conventional hybrid and pSIC-hybrid is smallest for the optimized hybrids, {\it i.e.} $\alpha_m$, as suggested by \eq{eq:hyb-pSIC}.
  }
  \label{fig:Epol_mixing}
\end{figure}

While Ref.~\onlinecite{GygBal89} justifies the assumption of $\vec{R}$-independence of $\Delta_\text{XC}^{\pm}$, it also suggests that the explicit $\vec{R}$-dependence of $\Delta^{\pm}_\text{XC}$ is to large extent determined by the static COHSEX approximation, and thus little is gained by employing full dynamic screening within the $GW$ approximation. A reasonable way to incorporate contributions from static screening of exchange in the total-energy functional is through the hybrid technique. Of course the fraction of EX in the hybrid functional, or in other words the magnitude of static screening, needs to be determined {\it a priori}.

\Eq{eq:final} describes a general formalism for correcting polaron self-interaction in any XC functional. Up to now, our discussion has been primarily concerned with calculations based on semilocal XC functionals such as PBE, but everything said still holds if one were to correct a hybrid-XC functional for polaron self-interaction. Bear in mind that for orbital-dependent functionals, such as hybrid-XC functionals, the chemical potentials $\mu_\text{hyb}^{\pm}$, correspond to the eigenvalues of a generalized KS Hamiltonian with a non-local self-consistent potential. Hence the XC derivative discontinuity $\Delta_\text{XC}^{\pm}$, is already incorporated in the chemical potentials,\cite{MorCohYan08} provided the hybrid functional is correctly parametrized. 

For a class of hybrid functionals, where PBE is combined with a fraction $0\leq \alpha \leq 1$, of EX, \Eq{eq:final} takes the forms
\begin{align}
\Delta E_{\text{pSIC-hyb}}^{\pm}[\vec{R};\alpha] = \Delta E_{\text{hyb}}[N_e;\vec{R};\alpha] \pm \Delta \mu_{\text{hyb}}^{\pm}[\vec{R};\alpha],
 \label{eq:finhyb}
\end{align}  
The energy of a polaron in configuration ${\vec{R}}$ relative to the undistorted crystal ${\vec{R_0}}$, can be estimated in two ways: either through the conventional approach using the functional $\Delta E_{\text{hyb}}[N_e\pm 1;\vec{R};\alpha]$, or through the pSIC approach using the functional $\Delta E_{\text{pSIC-hyb}}^{\pm}[\vec{R};\alpha]$. The terminology {\it conventional} versus pSIC approach will be used throughout the remainder of this paper. The detailed expressions can be found in \sect{sec:binding} of the Appendix. 

The difference between the conventional and the hybrid approach is illustrated in \fig{fig:Epol_mixing}. It shows polaron energies as a function of the hybrid mixing parameter $\alpha$, for two alkali halide compounds NaI and NaF, calculated using the conventional and the pSIC methods. Note that the polaron energies vary significantly with $\alpha$, when calculated using the conventional energy functional $\Delta E_{\text{hyb}}[\pm 1;\vec{R};\alpha_m]$, whereas they depend very weakly on $\alpha$, when calculated using the pSIC functional $\Delta E_{\text{pSIC-hyb}}^\pm[\vec{R};\alpha]$. The insensitivity of the pSIC-hybrid functional to parametrization is the main advantage of the pSIC method. It allows for accurate predictions of polaron energies with $\alpha$ set to zero, {\it i.e.} pSIC-DFT. In this way the computational cost of the calculations can be reduced significantly compared to hybrid functionals, and brought to be on par with simple DFT calculations with a minor impact on accuracy. 

\Fig{fig:Epol_mixing} shows that for a certain mixing parameter $\alpha_m$ the conventional and the pSIC methods are equivalent. This can be formally expressed as follows:
\begin{align}
\Delta E_{\text{pSIC-hyb}}^{\pm}[\vec{R};\alpha_m] = \Delta E_{\text{hyb}}[N_e\pm 1;\vec{R};\alpha_m].
\label{eq:hyb-pSIC}
\end{align}
This implies that at  $\alpha_m$, the total energy as a function of fractional hole-polaron charge has negligible curvature and the hybrid XC is thus polaron self-interaction free \cite{ErhKleAbe14}; in this sense, $\alpha_m$ can be considered the optimal mixing parameter. We will see in \sect{sec:results} that for the materials studied in this work, the choice of $\alpha$ that reproduces the DFT+$G_0W_0$ band gaps, nearly coincides with $\alpha_m$.

\subsection{Numerical implementation of energy and forces}

We now discuss the numerical implementation of the pSIC energy functional. Equations~(\ref{eq:final}) and (\ref{eq:finhyb}) comprise two terms, namely (i) the total energy in the reference charge state and (ii) the chemical potential in the reference charge state.
By invoking Koopman's theorem, the latter quantities can be calculated from the eigenvalue spectrum of the KS Hamiltonian in the case of DFT calculations, and the generalized KS Hamiltonian with a non-local self-consistent potential in the case of hybrid calculations. In particular, for electron polarons, $\mu^+$ correponds to the conduction band minimum (CBM) and for hole polarons, $\mu^-$ corresponds to the VBM. 

Structural relaxations and molecular dynamics simulations require the calculation of atomic forces. Equations~(\ref{eq:final}) and ~(\ref{eq:finhyb}) are, however, not directly suitable for evaluation of atomic forces via the Hellman-Feynman theorem, since $\mu_{\text {DFT}}^{\pm}$ is not a variational quantity. Explicit calculations of wave function derivatives with respect to ionic coordinates can be avoided by deducing $\mu_{\text {DFT}}^{\pm}$ via replacing \eq{eq:mu} by a finite difference formula such as
\begin{align}
  \mu_{\text {DFT}}^{\pm} =& \frac{1}{2\delta} \big( 4E_{\text{DFT}}[N_e\pm2\delta] - E_{\text{DFT}}[N_e] - E_{\text{DFT}}[N_e\pm\delta] \big)
  \label{eq:fvar}
\end{align}
where $\delta \ll 1$ and the dependence on $\vec{R}$ has been omitted for brevity. This relation, which is correct to the third order in $\delta$, in conjunction with \eq{eq:final} allows for the atomic forces to be calculated from a linear combination of three separate Hellmann-Feynman forces as follows: 
\begin{align}
\frac{\partial E_\text{pSIC}^\pm}{\partial\vec{R}} = & \left(1\mp\frac{1}{2\delta}\right) \frac{\partial E_{\text{DFT}}}{\partial\vec{R}}[N_e] \pm \nonumber \\ 
&\frac{1}{2\delta} \left( 4\frac{\partial E_\text{DFT}}{\partial\vec{R}}[N_e\pm2\delta]-\frac{\partial E_\text{DFT}}{\partial\vec{R}}[N_e\pm\delta] \right).
  \label{eq:fvar1}
\end{align}
The accuracy of this expression can be established by comparing the chemical potential calculated via \eq{eq:fvar}  to the one obtained from the eigenvalue spectrum. The numerical stability of the atomic forces from \eq{eq:fvar1} mainly depends on the size of the finite-difference increment $\delta$. We have found that a value of $\delta=0.025$ is sufficient to obtain accurate energies and converge forces to below 1\,meV/\AA.

We have implemented the pSIC method in the Vienna {\it ab initio} simulation package.
At each ionic step, the wave functions of three electronic configurations are converged independently, representing the electronic states $N_e$, $N_e+\delta$, and $N_e+2\delta$ for electron polarons, and $N_e$, $N_e-\delta$, and $N_e-2\delta$ for hole polarons. The energies and forces are collected from all replicas and combined according to Eqs.(~\ref{eq:fvar}) and (\ref{eq:fvar1}). It is apparent that in this parallel implementation, a pSIC-DFT calculation requires three times as much wall time as a standard DFT calculation.

\section{Results}
\label{sec:results}

\subsection{$V_K$-center formation in alkali halides}

Let us start this section by recapitulating the pSIC-DFT results for the $V_K$-center in NaI. It is apparent from \fig{fig:self_trapping_path}, that in contrast to conventional PBE calculations, the new pSIC-PBE functional predicts a stable $V_K$-center configuration, the geometry and energetics of which are in excellent agreement with the optimized hybrid functional. The bond length of the I$_2^-$ dimer that forms the core of the $V_K$-center is calculated to be 3.38\,\AA\ (3.23\,\AA) when using the pSIC-PBE (optimized hybrid) scheme while the energy gain due to self-trapping is found to be 0.267\,eV (0.345\,eV). 

In the following, we further evaluate the accuracy of the pSIC-PBE method by an extensive study of its predictions for the $V_K$-centers in alkali halides. These compounds constitute a diverse class of wide-gap insulators with band gaps ranging from 5 to 14\,eV and lattice constants from 4.0 to 7.3\,\AA, see \tab{tab:crystaldata}. We compare the results of pSIC-PBE calculations with conventional hybrid-XC calculations using two distinct parametrizations: (i) the PBE0 functional \cite{PerErnBur96}, which uses a fraction of 0.25 of EX, and (ii) a set of optimized hybrid functionals, which have been parametrized to reproduce PBE+$G_0W_0$ calculations. For these compounds, the fraction of EX used in the optimized hybrid calculations vary from 0.25 in LiI to 0.41 in NaF. Further details regarding the $G_0W_0$ calculations are provided in \sect{sect:GW_calculations} of the Appendix.

\begin{figure}
  \centering
  \includegraphics[scale=\myscale]{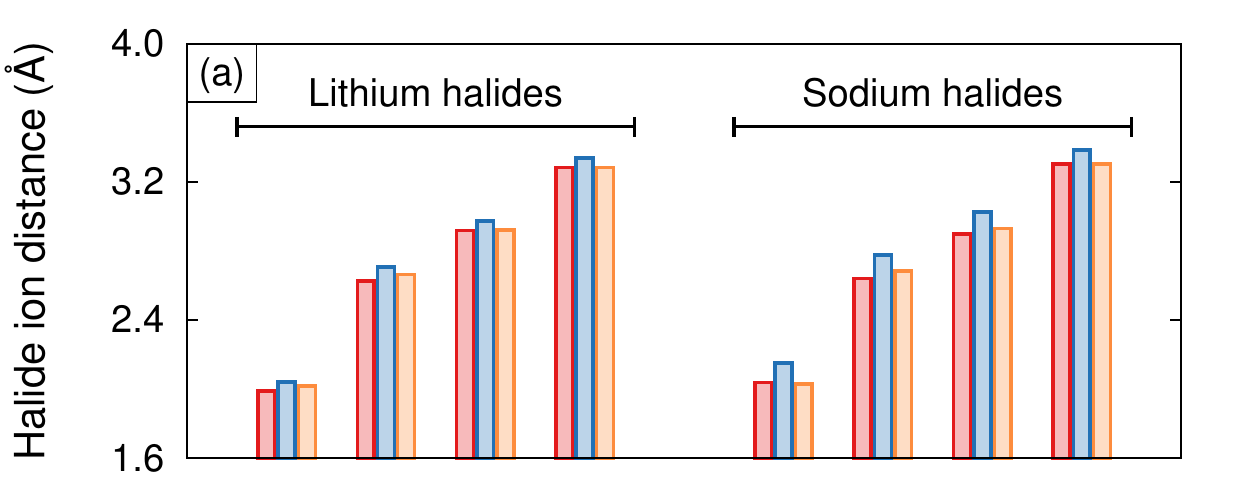}
  \includegraphics[scale=\myscale]{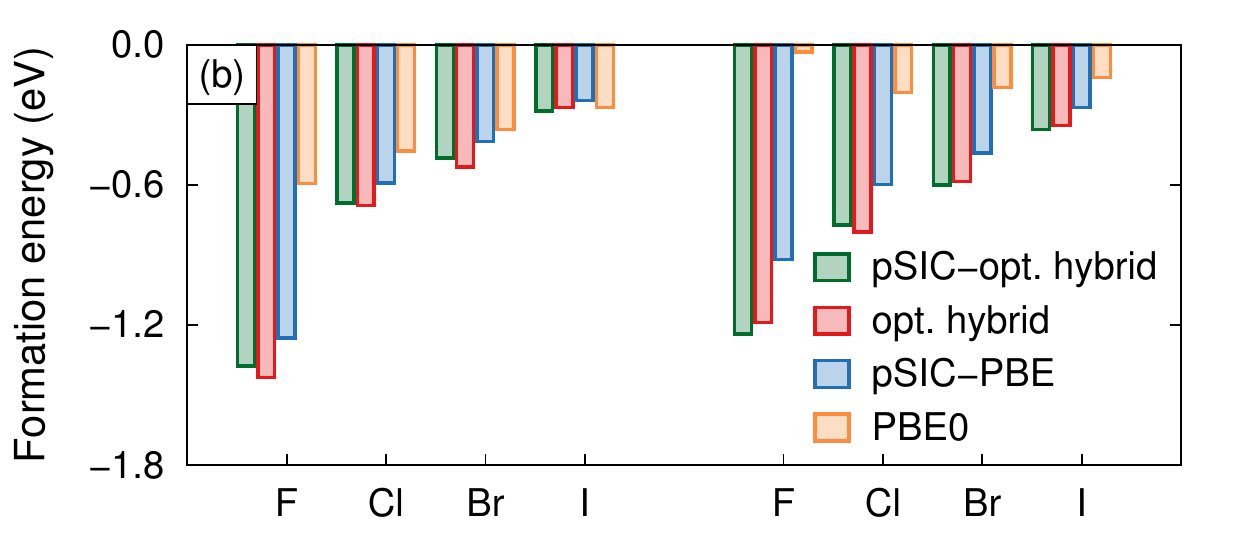}
  \caption{
    Comparison of (a) halogen--halogen separation at the core of the $V_K$-center and (b) $V_K$-center binding energies between pSIC-PBE and hybrid-DFT calculations.
  }
  \label{fig:comparison}
\end{figure}

The results of pSIC-PBE and hybrid calculations are compiled in \fig{fig:comparison}, which shows (a) the halogen--halogen separation at the core of the $V_K$-center and (b) the $V_K$-center binding energies. With regards to the halogen dimer separations, it is evident that the pSIC-PBE method yields excellent agreement with hybrid calculations with a typical deviation of less than 0.15\,\AA\ ($\lesssim\,5\%$). 

\Fig{fig:comparison}(b) shows that the polaron binding energies calculated within pSIC-PBE agree well with the results from optimized hybrid calculations. Overall, pSIC-PBE yields polaron binding energies that are 10 to 20\%\ smaller than those from optimized hybrids. On the other hand, for most materials considered here, especially those with larger band gaps, the PBE0 functional, in the conventional mode of calculation, predicts $V_K$-center binding energies that are much smaller than either pSIC-PBE or optimized hybrid calculations. This is consistent with a systematic underestimation of the band gaps of these wide-gap materials by PBE0, see \tab{tab:crystaldata}. As already discussed, further reducing the fraction of EX from 0.25 to 0, {\it i.e.} PBE, leads to an even larger underestimation of band gaps and, more dramatically, the failure to predict hole localization and polaron formation in all alkali halides. 

In summary, the potential energy landscapes of polarons, calculated within conventional hybrid schemes are highly dependent on the particular parameterization of the hybrid XC functional employed, {\it i.e.} the fraction of EX included in the calculations. The pronounced dependence of polaron binding energies on the hybrid parametrization is further illustrated for NaI and NaF in \fig{fig:Epol_mixing}, highlighting a major disadvantage of the conventional approach.

In contrast, the pSIC functional is quite insensitive to the fraction of EX chosen for the hybrid calculations. To illustrate this, we have performed pSIC-hybrid calculations of the polaron binding energies using \eq{eq:finhyb}. For each compound, calculations were carried out in the relaxed ionic configuration obtained from a conventional optimized hybrid calculation, using positively charged supercells. The results are compiled in \fig{fig:comparison}(b), where for every compound, pSIC-hybrid calculations are shown side by side with conventional hybrid results. \Fig{fig:comparison}(b) shows nearly perfect agreement between the pSIC and the conventional approaches for the optimized hybrid parametrization. This suggests that for the compounds in this study, the hybrid parametrizations chosen to reproduce PBE-$G_0W_0$ band gaps, yield nearly polaron-self-interaction-free XC functionals, for which \eq{eq:hyb-pSIC} suggests equality of the pSIC and the conventional approaches. However, a far more important conclusion can be reached when recollecting that the pSIC-DFT calculations of the polaron energies shown in \fig{fig:comparison}(b), are already within 10-20\% of the optimized pSIC-hybrid estimates: While the polaron binding energies are highly sensitive to hybrid parametrization within the conventional approach, this variation is an order of magnitude smaller in the pSIC approach. This is clearly illustrated for NaI and NaF in \fig{fig:Epol_mixing}, where the polaron binding energies as calculated within the pSIC and the conventional hybrid approaches are shown in the same graph. Note that again the two curves cross at $\alpha_m$, see \eq{eq:hyb-pSIC}, which is the optimized fraction of EX in the hybrid functional.

Finally, it is in order to discuss the sources of error in the pSIC-hybrid functional \eq{eq:finhyb}, which lead to the variation of $\Delta E_{\text{pSIC-hyb}}^\pm[\vec{R},\alpha]$ with the parameter $\alpha$. For this purpose, we focus on NaF, for which pSIC-PBE has the largest discrepancy with optimized hybrid, of any of the compounds considered in this paper. For NaF, pSIC-PBE predicts a polaron binding energy of 0.92 eV versus 1.19 eV for optimized hybrid in the conventional approach, and 1.24 eV for optimized hybrid in the pSIC approach.

The relatively large error (in excess of 20\%) of the polaron binding energy in NaF, when calculated within pSIC-PBE can be related to the individual error of each term in \eq{eq:finhyb}. For the case of hole polarons in NaF, \eq{eq:finhyb} consists of two contributions: (i) neutral ground state energy, and (ii) the energy of the VBM. Both quantities are calculated as the difference between the polaron configuration and the perfect lattice. The first term is calculated within PBE (optimized hybrid) to be $-2.99$ ($-3.06$), while the second term is $3.85$ ($4.30$). The relative ground-state energies differ by less than $3\%$ between PBE and optimized hybrid, while the error in the relative VBM levels from PBE is no more than $10\%$. The relatively large error in excess of 20\%, in the final polaron binding energy is caused by the opposite signs of the two contributions.

\subsection{$V_K$-center migration in NaI}

\begin{table}
  \caption{
    Activation barriers for $V_K$-center migration in NaI in units of eV from calculation and experiment \cite{PopMur72}.
  }
  \centering
  \begin{tabular}{c c c c}
    \hline\hline
    Rotation angle & Optimized hybrid & pSIC-PBE & Experiment \\ [2pt]
    \hline
    $60^{\circ}$  & 0.23 & 0.20  &  0.2         \\
    $90^{\circ}$  & 0.28 & 0.24  &  $>60^{\circ}$ \\
    $120^{\circ}$ & 0.23 & 0.20  &  0.2         \\
    $180^{\circ}$ & 0.22 & 0.19  &              \\[2pt]
    \hline \hline
  \end{tabular}
  \label{tab:migration}
\end{table}

Thanks to its computational efficiency, the pSIC-DFT method provides an unprecedented potential for studying polaron dynamics. To demonstrate this aspect, we also benchmark calculated polaron migration barriers with measured diffusivities in NaI. By symmetry, the migration of an I$_2^-$ dimer to a nearest-neighbor site in the rocksalt lattice can involve rotations through $60^{\circ}$, $90^{\circ}$, $120^{\circ}$ or $180^{\circ}$, the last one corresponding to a pure translation. The polaron diffusivity in alkali halides can be accurately measured in laser pump-probe experiments \cite{PopMur72}. To this end, $V_K$-centers are created and aligned along a particular direction (``polarized'') using a pump laser. Subsequently, the gradual loss of polarization over time is followed using a probe laser, from which rotational barriers can be inferred.

Experiments cannot distinguish between $60^{\circ}$ and $120^{\circ}$ rotations and are insensitive to migration via translation. In NaI at 50\,K, one observes almost exclusively $60^{\circ}/120^{\circ}$ rotations \cite{PopMur72}, indicative of the fact that their activation barriers are noticeably smaller than for $90^{\circ}$ rotations. It was similarly shown that $60^{\circ}/120^{\circ}$ rotations dominate in RbI \cite{DelStoYus77,ShlKanHei92}. \tab{tab:migration} shows activation barriers calculated within the pSIC-PBE and optimized hybrid approaches using the climbing image nudged elastic band method \cite{HenUbeJon00}. Both methods agree very well with each other and with experimental data. 

\begin{figure}
  \centering
  \includegraphics[width=0.65\linewidth]{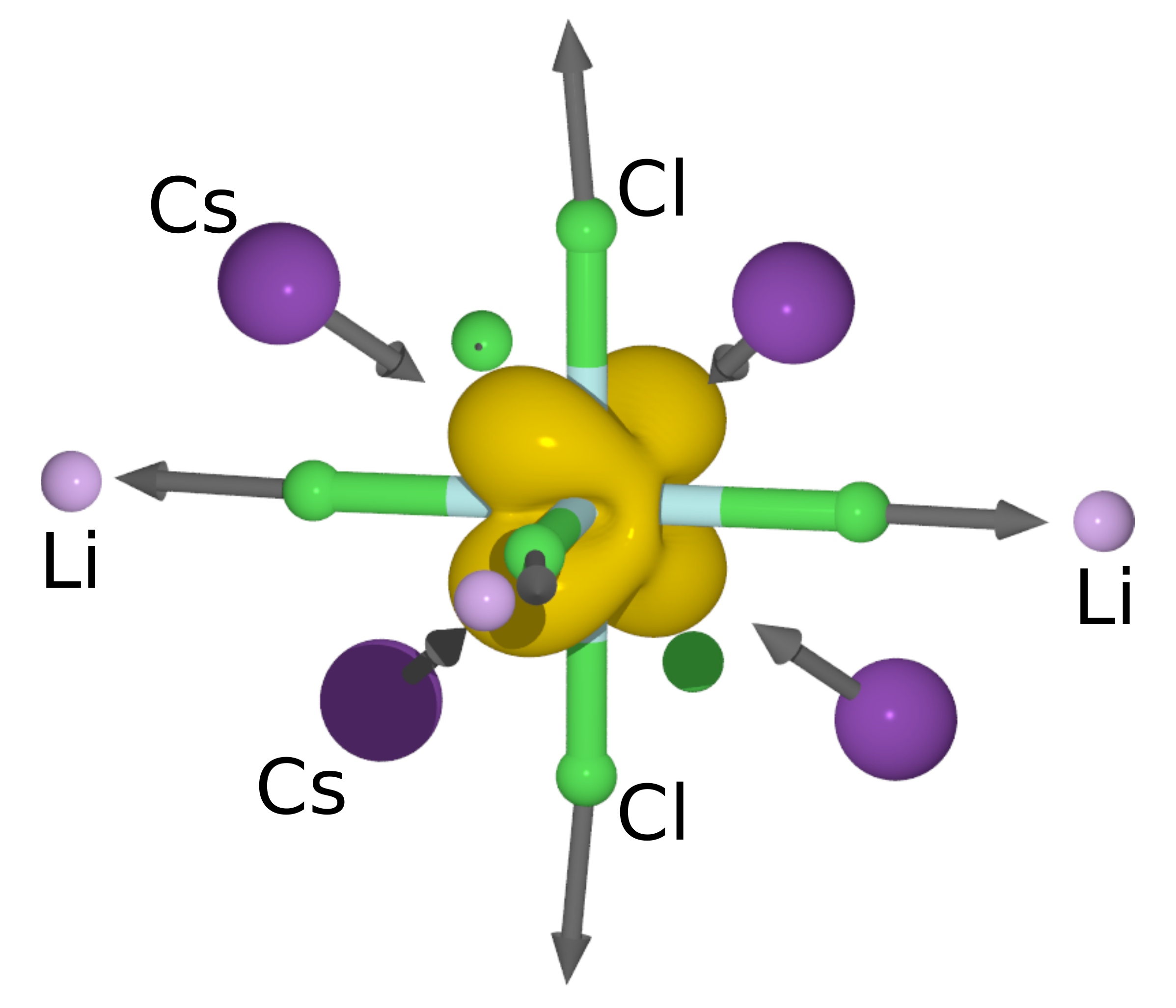}
  \caption{
    Illustration of the electron polaron in CLYC as obtained from pSIC calculations. The arrows indicate the relaxation pattern. The yellow isosurface is centered at an yttrium atom.
  }
  \label{fig:CLYC}
\end{figure}

\begin{figure}
  \centering
  \includegraphics[scale=\myscale]{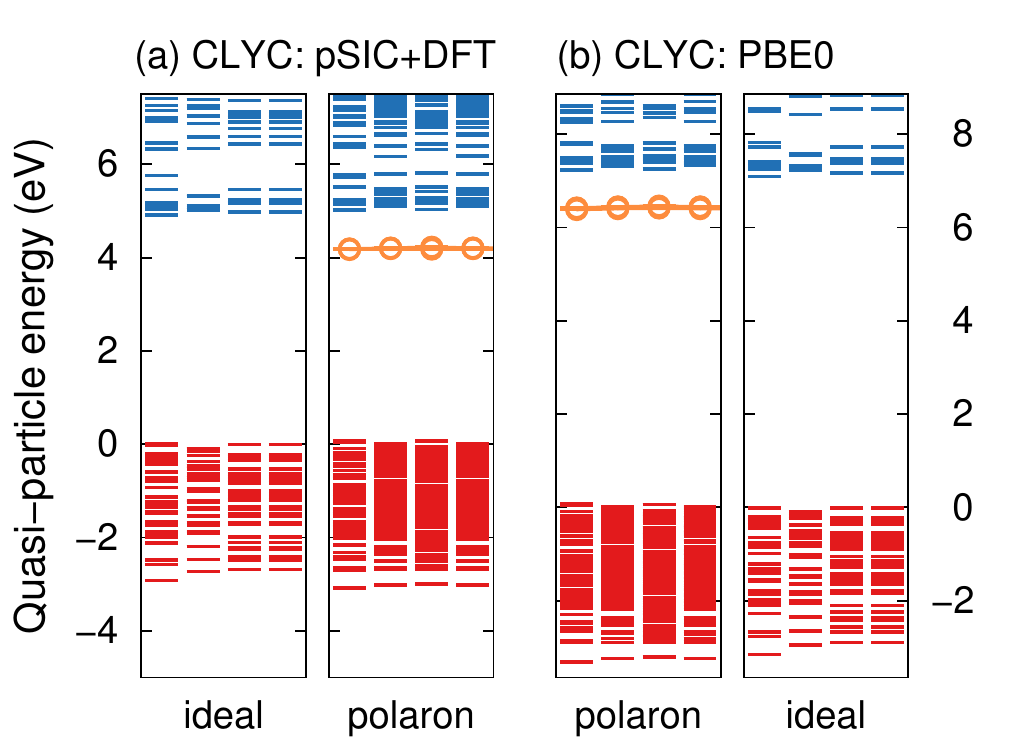}
  \caption{
   Energy levels for the ideal and polaronic configuration calculated using (a) DFT and (b) PBE0 for CLYC. Red and blue lines represent occupied and unoccupied levels, respectively. The polaronic level is shown by orange lines and open circles.
  }
  \label{fig:CLYC_bnd}
\end{figure}

\subsection{Electron polarons in CLYC}

Lastly, we demonstrate the accuracy of the pSIC method for electron polarons. For this purpose, we consider Cs$_2$LiYCl$_6$ (CLYC), which is one of the most thoroughly investigated compounds in the elpasolite family due to its potential as a neutron detector. Electron polarons have been found in this system both experimentally \cite{PawSpa97} and theoretically \cite{BisDu12} using PBE0 calculations. Its crystal structure (double-perovskite) has cubic symmetry with ten atoms in the primitive cell. Here, electron polaron calculations were performed using 80-atom supercells with a $2\times2\times2$ $\vec{k}$-point sampling. The calculated band gap within PBE0 is 7.1\,eV, in close agreement with experiment (7.5\,eV) \cite{BesDorEij04}. Hence PBE0 and optimized hybrid yield nearly identical results for this compound. By comparison, DFT-PBE yields a much smaller band gap of 5.0\,eV. We find in accord with Ref.~\onlinecite{BisDu12} that the electron polaron localizes on a Y-site and the polaron level has strong $d$-character as shown in \fig{fig:CLYC}. Both pSIC-PBE and PBE0 predict a stretching of Y-Cl bonds by about 0.1\,\AA. The binding energy of the electron polaron is 0.25\,eV (0.32\,eV) according to pSIC-PBE (PBE0). The PBE0 value for the polaron binding energy includes a significant contribution from image charge corrections of about 0.1\,eV due to the small size of the supercells (80 atoms) that was computationally affordable at this time. (Note that the hitherto most recent calculation for the electron polaron in this system was performed in a 40-atom supercell \cite{BisDu12}). In fact, performing pSIC-PBE0 calculation of the polaron energy in the ionic confguration obtained from the conventional PBE0 approach, yields a polaron binding energy of about 0.28 eV, in very good agreement with the pSIC-PBE result. This demonstrates that pSIC is also capable of predicting {\em electron} polaron geometries and binding energies in close agreement with hybrid calculations.

In order to illustrate that this agreement is not fortuitous, we show in \fig{fig:CLYC_bnd} the quasi-particle spectra of CLYC calculated within DFT, as well as PBE0 for the ideal and polaron configurations in their charge-neutral states. Note that the scissor shift is independent of configuration and that the calculated position of the localized electron level relative to the conduction-band minimum within DFT agrees well with PBE0.

\section{Conclusions and outlook}
\label{sec:discussion}

In this paper we have introduced a self-interaction correction technique, the so-called pSIC method, that corrects the DFT failure to predict stable polarons in insulators. The pSIC-DFT method is parameter-free and easy to implement in existing electronic structure codes. We have validated the new methodology by benchmarking pSIC-PBE results to hybrid-DFT functionals for both electron and hole polaron formation and transport in a number of materials. Thereby, we have shown that (i) polaron trapping energies depend sensitively on the parametrization of the hybrid functionals, (ii) the parameter-free pSIC-PBE method exhibits similar accuracy to hybrid functionals that are laboriously parametrized to fit $GW$ band gaps on a case by case basis, and (iii) is an order of magnitude faster. This presents a breakthrough in modeling the structure and dynamics of polarons in insulators from first principles. 

From a fundamental physics perspective, the pSIC method emerges as a result of the decoupling of the two main problematic features of DFT for band insulators: (i) severe underestimation of band gaps, and (ii) absence of localized polarons. The counterintuitive aspect of such a decoupling is that it is well-known that polaron stability is greatly enhanced for large-gap systems. Hence significant correlation between the band gap of an insulator and its polaron binding energy should be expected. Based on this, it is then logical to argue that the lack of stable polarons in insulators within the DFT must stem from its band gap error. In fact, the two errors are unrelated as shown by the following arguments: (i) for the exact XC functional, the electron addition/removal energies can be extracted from differential quantities, {\it i.e.} chemical potentials, (ii) for most band insulators, there is a charge state, usually the closed-shell neutral charge state, for which DFT can predict accurate ground-state energies as well as valence and conduction band structures, and (iii) while LDA and GGA severely underestimate the band gaps of insulators, they accurately predict band edge variations with configurational changes.  Noting that polaron binding does sensitively depend on the variation of the ground-state energy and band edges with configuration rather than the absolute value of the scissor shift, it follows that polaron binding is independent of the absolute value of the band gap error.   

In order to rationalize the success of the pSIC approach, we have applied it to correct for the pSI in hybrid XC functionals. In this process, we have shown that in band insulators, hybrid functionals tuned to reproduce the DFT+$G_0W_0$ band gaps are practically polaron self-interaction error free, and thus constitute the optimal parametrization. Our key finding however, is that calculations of polaron energies within the conventional approach, {\it i.e.} using charged supercells, are very sensitive to the fraction $\alpha$ of EX included in the hybrid functional. On the other hand, the pSIC-hybrid method is almost insensitive to $\alpha$, and little compromise on accuracy will result from drastically simplifying the calculations by setting $\alpha = 0$. On this basis, the parameter-free pSIC-PBE method emerges. 

The pSIC formalism can be applied to more general systems than has been presented in this paper. For the sake of clarity, we have intentionally limited the present scope to intrinsic polaron self-trapping in crystalline insulators. We have shown that in these systems, the neutral charge state is described accurately within PBE (or more generally DFT). Hence this reference state has been used to calculate the energy of charge excitations in the presence of lattice distortions using \eq{eq:hyb-pSIC}. More generally, the pSIC formalism can be applied to both defects in insulating solids, as well as molecular systems whenever structural rearrangements are driven by addition/removal of electron or holes in frontier orbitals. The pSIC method allows any supercell calculation containing $N_e$ electrons, whether periodic or not, to be performed in any of the three charge states $N_e$, $N_e+1$, or $N_e-1$. We have demonstrated in this paper that in the case of perfect crystalline insulators, there is large variability with charge state with respect to the quality of the KS potentials produced from PBE. By choosing to perform the calculations in the charge state with the highest quality KS potential, the pSIC method enables dramatic enhancements of the accuracy of DFT XC functionals, for prediction of potential energy landscapes of polarons in these systems. 

The subject of defects in insulators and semiconductors is vast and complex, and of huge interest to many applications. Recently, the effect of $GW$ corrections on the formation energies of charged defects in semiconductors was examined \cite{RinJanSch09, LanZun10}. In these works, structural relaxations were performed using standard DFT, only to keep the computational costs down to a manageable level. 
It will be the subject of future to establish, if the pSIC method can also be used to obtain accurate {\em defect} configurations and energies at a reasonable computational cost.

\begin{acknowledgments}
We thank Michael Surh at LLNL for very helpful discussions. Lawrence Livermore National Laboratory is operated by Lawrence Livermore National Sec\-urity, LLC, for the U.S. DOE-NNSA under Contract DE-AC52-07NA27344. Funding for this work was received from the NA-22 agency. P.E. acknowledges funding from the Knut and Alice Wallenberg Foundation and the {\em Area of Advance -- Materials Science} at Chalmers. Com\-puter time allocations by the Swedish National Infrastructure for Computing at NSC (Link\"oping) and C3SE (Gothenburg) are gratefully acknowledged.
\end{acknowledgments}

\appendix*

\section{Computational details}
\label{sec:compdet}

\subsection{Total-energy calculations}

Calculations were carried out using the projector augmented wave method \cite{Blo94, *KreJou99} as implemented in the Vienna {\it ab initio} simulation package \cite{KreHaf93, *KreFur96b}, using the supplied standard plane wave energy cutoffs. Three classes of XC functionals are juxtaposed in this work: (i) PBE, (ii) hybrid-DFT, and (iii) pSIC-DFT. The two latter functionals also build upon the PBE XC functional \cite{PerBurErn96}. For the alkali halides we employed 216-atom supercells and the Brillouin zone was sampled using the $\Gamma$-point only. For the CLYC system we employed 80-atom supercells with a $2\times2\times2$ $\vec{k}$-point sampling.

\subsection{Polaron binding energies}
\label{sec:binding}

Polaron formation (or binding) energies are computed in the pSIC formalism simply as 
\begin{align}
  \Delta E_f[\boldsymbol{R}_{\text{pol}}] = E^{\pm}_\text{pSIC}[\boldsymbol{R}_{\text{pol}}]  - E^{\pm}_\text{pSIC}[\boldsymbol{R}_{\text{ideal}}] ,
\end{align}
where $E^{\pm}_\text{pSIC}$ is defined in \eq{eq:final0}.

{\em Conventional} hybrid calculations determine polaron binding energies through energy differences between {\em charged} and {\em neutral} supercells (see Refs.~\onlinecite{ErhKleAbe14} and \onlinecite{LanZun08}) according to
\begin{align}
  \Delta E_f &= E_{\text{PBE0}}[\pm 1, \boldsymbol{R}_{\text{pol}}] - E_{\text{PBE0}}[0, \boldsymbol{R}_{\text{ideal}}] \pm \varepsilon^{\text{KS}}_\pm,
  \label{eq:standardform}
\end{align}
which requires image charge corrections to be applied. This causes further uncertainties as there is no exact correction procedure \cite{KomPas11}. Note that in contrast the pSIC method relies on the (electron) chemical potentials in the neutral reference state for obtaining polaron energies, and thus is not subject to image charge corrections. In the case of the present hybrid calculations, the spurious image charge interaction is taken into account via the modified Makov-Payne correction of Lany and Zunger\cite{MakPay95, LanZun08}
\begin{align}
  \Delta E_{corr} = \frac{2}{3}\frac{M}{2\epsilon L},
  \label{eq:MP}
\end{align}
where $M$ is the Madelung constant and $L$ is the lattice constant of the supercell. The dielectric constants used in \eq{eq:MP} is given by
\begin{align}
  \epsilon = \epsilon_\infty+ \epsilon_\text{ion},
\end{align}
where the first and second terms denote the electronic and ionic contribution, respectively. We approximately include local field effects in $\epsilon_\infty$ for the hybrid functionals by directly applying the difference between DFT dielectric constants with and without local field effects. Computational input parameters and calculated properties such as lattice constants, band gaps, and dielectric constants are listed in \tab{tab:crystaldata}. We finally note that \eq{eq:standardform} is equivalent to the relaxation energy $\Delta E_{\text{hyb}}[N_e\pm 1;\vec{R};\alpha]$ in the infinite dilution limit \cite{ErhKleAbe14}.

\begingroup
\begin{table}
\centering
\caption{\label{tab:crystaldata}
  Lattice constants, band gaps, and dielectric constants for the alkali halides and CLYC. The dielectric constants for the latter are slightly anisotropic and has been averaged over the three principal axes.
}
\begin{ruledtabular}
\begin{tabular}{lYYYYYY}
 & & \multicolumn{3}{c}{Band gap (eV) } \\
\cline{3-5}
&   \multicolumn{1}{c}{$a$ (\AA{})}  &
    \multicolumn{1}{c}{PBE}  &
    \multicolumn{1}{c}{PBE0}  &
    \multicolumn{1}{c}{$G_0W_0$}  &
    \multicolumn{1}{c}{$\alpha_{\text{EX}}$} &
    \multicolumn{1}{c}{$\epsilon$}
\\
\hline
LiF  & 4.065  &  8.84  &  11.85  &  13.24   &  0.37 &  10.55  \\
LiCl & 5.090  &  6.46  &   8.57  &   9.08   &  0.31 &   9.91  \\
LiBr & 5.452  &  5.07  &   6.97  &   7.30   &  0.29 &  14.45  \\
LiI  & 5.984  &  4.32  &   5.95  &   5.98   &  0.25 &  19.43  \\[3pt]
NaF  & 4.537  &  6.64  &   9.60  &  11.60   &  0.41 &   4.25  \\
NaCl & 5.578  &  5.22  &   7.33  &   8.42   &  0.38 &   5.66  \\
NaBr & 5.916  &  4.24  &   6.15  &   6.94   &  0.35 &   5.89  \\
NaI  & 6.407  &  3.72  &   5.39  &   5.80   &  0.31 &   7.07  \\[3pt]
CLYC & 10.485 &  4.89  &   7.12  &   \multicolumn{1}{r}{$-$}    &  0.25 & 9.85
\end{tabular}
\end{ruledtabular}
\end{table}
\endgroup

\subsection{$G_0W_0$ calculations}
\label{sect:GW_calculations}

Quasiparticle $G_0W_0$ energies \cite{Hed65, *HedLun70, *AulJonWil00} were calculated for the alkali halides in the rocksalt crystal structure on top of PBE using $\Gamma$-centered $6\times 6\times 6$ Monkhorst-Pack grids as implemented in the Vienna ab initio simulation package\cite{ShiKre06, *ShiKre07, *FucFurBec07}. We included 192 bands to converge the dielectric function used in the screened interaction $W_0$ together with a frequency grid with 48 points. 

Quasiparticle $G_0W_0$ calculations were also performed for the self-trapped hole polaron in NaI. They employed a 64-atom supercell and calculated $G_0W_0$ corrections on top of the PBE0 hybrid functional using 1080 bands for the dielectric function together with 24 points in the frequency grid.

\end{document}